\documentclass{article}
\usepackage{graphicx}  
\usepackage{amsmath}   
\usepackage[compress]{cite}
\usepackage{amssymb}   
\usepackage{bm} 
\usepackage{dcolumn}
\usepackage{color}
\usepackage{mathrsfs}
\usepackage{amsfonts}
\usepackage{varioref}
\usepackage{textcomp}
\RequirePackage[colorlinks,citecolor=blue,urlcolor=magenta,linkcolor=blue]{hyperref}
\allowdisplaybreaks
\def\gr{general relativity}

\labelformat{section}{Section #1} 
\labelformat{subsection}{Section #1} 
\labelformat{subsubsection}{Section #1}
\labelformat{subsubsubsection}{Section #1}
\labelformat{equation}{~(#1)} 
\labelformat{figure}{Fig.~#1} 
\labelformat{subfigure}{Fig.~\thefigure#1} 
\labelformat{table}{Tab.~#1} 
\labelformat{appendix}{Appendix #1}

\def\no{\nonumber}

\def\p{\partial}

\def\na{\nabla}
\def\tna{\tilde{\nabla}}

\def\k{\kappa}

\def\lie{\pounds_{\xi}}
\def\t{\widetilde}
\def\tlie{\pounds_{\t{\xi}}}

\def\be{\begin{equation}}
\def\ee{\end{equation}}
\def\ba{\begin{align}}
\def\ea{\end{align}}

\def\mg{\sqrt{-g}}

\def\tmg{\sqrt{-\tilde g}}

\def\A{\mathcal{A}}
\def\R{\mathcal{R}}
\def\ST{scalar-tensor }
\title{\bf Scalar-tensor theories of gravity from a thermodynamic viewpoint}

\author{Krishnakanta
Bhattacharya\footnote{krishnakanta@dubai.bits-pilani.ac.in}$~^{1}$ and Sumanta Chakraborty\footnote{tpsc@iacs.res.in}$~^{2}$
\\
$^{1}${\small{Department of General Science}}\\
{\small{BITS Pilani, Dubai Campus, International Academic City, Dubai, United Arab Emirates}}\\
$^{2}${\small{School of Physical Sciences}}\\
{\small{Indian Association for the Cultivation of Science, Kolkata-700032, India}}}
\begin{document}
  
\maketitle
\begin{abstract}
In any diffeomorphism invariant theory of gravity, one can define a Noether charge arising from the invariance of the Lagrangian under diffeomorphisms. We have determined the Noether charge for scalar-tensor theories of gravity, in which case the gravity is mediated by the metric tensor as well as by a scalar degree of freedom. In particular, we demonstrate that the total Noether charge within an appropriate spatial volume can be related to the heat content of the boundary surface. For static spacetimes, in these theories, there exist an ``equipartition" between properly defined bulk and surface degrees of freedom. While the dynamical evolution of spacetime in these theories of scalar-tensor gravity arises due to the departure from the equipartition regime. These results demonstrate that thermodynamical interpretations for gravitational theories transcend Einstein and Lovelock theories of gravity, holding true for theories with additional scalar degrees of freedom as well. Moreover, they hold in both the Jordan and the Einstein frames. However, it turns out that there are two dynamically equivalent representations of the scalar-tensor theory in the Jordan frame, differing by total derivatives in the action, which are thermodynamically inequivalent. This depicts the importance of having a thermodynamic description, which can be used in distinguishing various dynamically equivalent representations of gravity theories beyond Einstein.
\end{abstract}
\section{Introduction}

The seminal works of the seventies first provided us with the insight that black holes are thermodynamic objects as one can associate them with temperature \cite{Hawking:1975vcx, Hawking:1976de}, entropy \cite{Bekenstein:1973ur, Bekenstein:1974ax}, and thermodynamic laws \cite{Bardeen:1973gs}. However, it was later realized that this tantalizing thermodynamic connection is more robust as one can associate temperature to a more general class of null surfaces \cite{Davies:1976ei, Unruh:1976db}, suggesting that the thermodynamic properties are not restricted to mere black holes, rather they have a wider range of applicability \cite{Wald:1999vt, Padmanabhan:2003gd}. Further works in this direction provided us with an alternative viewpoint towards understanding gravitation as an ``emergent phenomenon'', like thermodynamics or fluid mechanics, where the macroscopic phenomenon emerges from the dynamics of microscopic atoms \cite{Sakharov:1967pk, Jacobson:1995ab, Volovik:2000ua, Padmanabhan:2009vy}. Subsequently, it was found that this thermodynamic interpretation for gravitational dynamics holds for Lovelock class of theories as well. The crucial results in this regard are the following: (i) The gravitational actions can be decomposed into ``bulk'' and ``surface'' parts with a specific ``holographic relation'' between the two \cite{Padmanabhan:2004fq, Mukhopadhyay:2006vu, gravitation, Kolekar:2010dm, Kolekar:2011bb, Chakraborty:2016yna}; (ii) The dynamical equations of gravitation takes the form of thermodynamic identities upon projecting on the horizon \cite{Padmanabhan:2002sha, Cai:2005ra, Paranjape:2006ca, Akbar:2006er, Padmanabhan:2006ag, Kothawala:2009kc, Chakraborty:2015aja, Chakraborty:2015wma}, and most importantly these results hold for any generic null surface, not just black hole horizon \cite{Chakraborty:2015hna}; (iii) Gravitational field equations can be obtained from the thermodynamic extremization principle \cite{Padmanabhan:2007en, Padmanabhan:2007xy, Chakraborty:2015hna}; (iv) Gravitational actions can be interpreted as the free energy of the spacetime \cite{Gibbons:1976ue, Padmanabhan:2004fq, Padmanabhan:2009kr, Kolekar:2011bb}; (v) Gravitational field equations take the form of Navier-Stokes equation of fluid dynamics on a null surface \cite{DAMOUR, Gourgoulhon:2005ng, Padmanabhan:2010rp, Chakraborty:2015hna}, (vi) It has been shown that for every solution of incompressible  Navier-Stokes equation in ($p+1$) dimensions, one can associate a unique ``dual'' solution of Einstein's equation in vacuum in ($p+2$) dimensions \cite{Bredberg:2011jq}.

Along these lines, another pair of interesting results exists, where it has been found that --- (a) the total Noether charge contained within a three-volume $\R$ can be interpreted as the heat content of the surface $\p \R$ that encloses the volume; (b) it also allows to obtain the ``equipartition'' between suitably defined surface and bulk degrees of freedom in a static spacetime and any departure from equipartition results into the time evolution of the spacetime \cite{Padmanabhan:2013nxa, Chakraborty:2014rga, Chakraborty:2018qew}. The above thermodynamic interpretation of the gravitational field equations hold in general relativity, Lovelock theories of gravity as well as in Einstein-Cartan theories. For general relativity the above results can be summarized using the following relation \cite{Padmanabhan:2013nxa}:
\begin{align}
\int_{\R}d^3x h_{ab}\lie P^{ab} =\frac{1}{4}\k_{\rm B} T_{\rm avg} (n_{\rm bulk}-n_{\rm sur})~, 
\label{equipartgen}
\end{align}
where $h_{ab}$ is the induced metric on a spacelike surface, and $P^{ab}$ is the conjugate momentum of $h_{ab}$, with $\xi^a$ being the diffeomorphism vector field. Furthermore, $T_{\rm avg}$ is the average Unruh-Davies temperature of the boundary surface, and $n_{\rm bulk}$ and $n_{\rm sur}$ are the naturally defined bulk and surface degrees of freedom, respectively. The above result suggests that for a static spacetime, where $\xi^a$ becomes a timelike Killing vector, one obtains the equipartition relation $n_{\rm bulk}=n_{\rm sur}$. More importantly, the equation also implies that the departure from the equipartition ($n_{\rm bulk}\neq n_{\rm sur}$) is the mechanism that results into the time-evolution of the spacetime. Thus, Eq.\ref{equipartgen} provides us with a thermodynamic interpretation for the time-evolution of the spacetime, an insight which is not there in the geometrical version of the dynamical equations. For example, Einstein's equations in the form $G_{ab}=8\pi T_{ab}$, do not tell us why a spacetime should be time-dependent or not. 

Given the elegant description provided by Eq.\ref{equipartgen}, and the fact that it holds in Lovelock \cite{Chakraborty:2014rga} as well as Einstein-Cartan theories \cite{Chakraborty:2018qew}, besides \gr, it is important to study if such a relation exists in a more broader class of theories, namely the scalar-tensor theories of gravity. The scalar-tensor theory is one of the most popular among various alternative theories of gravity, and also arises in the low-energy effective actions of the string theory, where the spin-2 graviton couples with a spin-0 partner, known as the dilaton \cite{Buchbinder:1992rb, Damour:2002mi, Polchinski:1998rq}. In addition, it also follows that the $f(R)$ theories of gravity can be represented as a scalar-tensor theory \cite{Sotiriou:2008rp, DeFelice:2010aj, Nojiri:2010wj, FUJII}, and hence scalar-tensor theories arise in various contexts, starting from low-energy effective action of string theory to higher curvature theories of gravity. In the scalar-tensor theory, besides gravity, we have an additional degree of freedom, namely the scalar field $\phi$, which appears in the action via its non-minimal coupling with the Ricci scalar. This non-minimal coupling affects the dynamics and thermodynamical properties of this theory in a non-trivial manner.

In recent years, many of these results, which manifest a thermodynamic interpretation to the gravitational dynamics, have been obtained for scalar-tensor theories of gravity as well: (i) the decomposition of the gravitational action into bulk and surface terms \cite{Bhattacharya:2017pqc}; (ii) interpreting the action as the free energy of the spacetime \cite{Bhattacharya:2017pqc}; (iii) obtaining thermodynamical laws \cite{Koga:1998un, Bhattacharya:2018xlq, Bhattacharya:2022mnb, Bhattacharya:2021lgk}; (iv) deriving Navier-Stokes equation from field equations of scalar-tensor theory \cite{Bhattacharya:2020wdl}, and finally (v) obtaining gravitational field equations as thermodynamic identity \cite{Dey:2021rke}. Therefore, it is worth investigating whether the findings presented above, namely the interpretation of Noether charge as the heat content of the surface and the holographic equipartition relation, is also there in the scalar-tensor theories of gravity. In this connection it is worth pointing out that, due to the presence of the non-minimal coupling, the area law of entropy does not hold for \ST theories of gravity \cite{Kang:1996rj, Jacobson:1993vj, Visser:1993nu}, and hence it is not obvious that these results will naturally translate to these theories. 

Furthermore, the \ST theories of gravity are described in two frames which are conformally connected. The non-minimal coupling is present in the Jordan frame, while in the other frame, known as the Einstein frame, there is no non-minimal coupling. The physical equivalence of these two frames has been a matter of debate for ages \cite{Faraoni:1999hp, Faraoni:2022gry, ALAL,Faraoni:1998qx, Faraoni:2010yi, Faraoni:2006fx, Saltas:2010ga, Capozziello:2010sc, Padilla:2012ze, Jacobson:1993pf, Deser:2006gt, Kamenshchik:2014waa, Banerjee:2016lco, Chakraborty:2016ydo, Chakraborty:2016gpg, Pandey:2016unk, Ruf:2017xon, Karam:2017zno, Bahamonde:2017kbs, Karam:2018squ, Chakraborty:2023kel, Bhattacharya:2023ycc,Capozziello:2006dj,Bahamonde:2016wmz,Ohta:2017trn,Ohta:2018sze} and yet there is no universal consensus. Therefore, it will be interesting to see whether the thermodynamic parameters (like Noether charge) and other physical parameters (like the bulk and the surface degrees of freedom), which we obtain along the course of analysis, are equivalent in the two frames.

The paper is organized as follows: In \ref{reviewst} we provide a brief review of scalar-tensor theories of gravity and introuce the Jordan and Einstein frames. In addition, we also discuss the Noether current in each of these frames. Using these expressions for Noether current, in \ref{heatcontent}, we showed that the Noether charge can be interpreted as the surface heat content, and discuss the situation in each of the frames separately. Finally, the connection between the time evolution of the spacetime and holographic equipartition for both the frames have been presented in \ref{equipartition}. In \ref{concl}, we provide the concluding remarks for our analysis. Several supplementary discussions have been presented in the appendices. 

\emph{Notations and conventions:} We use mostly positive signature convention. Therefore, the four dimensional Minkowski metric reads $\textrm{diag.}(-1,+1,+1,+1)$. Throughout the paper, we set the fundamental constants such that $c=1=G$. The four-dimensional spacetime indices are denoted by Roman letters, e.g., $a,b,c,\cdots$. 

\section{Brief review of scalar-tensor theory of gravity} \label{reviewst}

In this section, we wish to review briefly the simplest sub-class of scalar-tensor theory of gravity, namely the Brans-Dicke theory of gravity and its generalization. Introduction of a scalar field to describe gravity dates back to the work of G. Nordstr$\ddot{\textrm{o}}$m, which was taken over by general relativity, describing gravitation in terms of the metric tensor, and subsequently ruled out by various experiments \cite{NORDSTROM}. Later, the Kaluza-Klien theories of gravitation \cite{KALUZA,KLIEN}, which acted as a precursor to the string theory, involves the metric, a vector field describing electromagnetism, and an additional scalar degree of freedom. Later, through Jordan and then by Brans and Dicke, such an extra scalar gives rise to the ``Brans-Dicke'' theory of gravitation \cite{Brans:1961sx}, acting as the precursor of all the scalar-tensor theories of gravity currently in use, including the Horndeski and the DHOST theories. Unlike GR, where the gravitation is described in terms of the metric tensor alone, in \ST gravity, the gravitation appears due to the interplay of the metric tensor as well as the scalar field. In what follows, we will review the action of the \ST theory of gravity and the associated dynamical equations. In particular, we will present our results in both the Jordan and the Einsten frames of reference and shall discuss their one-to-one correspondence. 

\subsection{Actions in the two frames and associated dynamical equations}\label{action_eq}

The Brans-Dicke theory of gravity is originally written in the Jordan frame with non-minimal coupling between the scalar field and metric. Simple generalization of the same yields the following action,
\begin{eqnarray}
&&\mathcal{A}_{\rm J}=\A^{\rm (grav)}[\boldsymbol{g},\phi]+\A^{\rm(mat)}[\boldsymbol{g},\psi]
=\int_{\mathcal{V}} d^4x\sqrt{-g}\Big[L^{\rm (grav)}(\boldsymbol{g},\phi)+L^{\rm(mat)}(\boldsymbol{g},\psi)\Big]~,
\no 
\\
&&\qquad =\int_{\mathcal{V}} d^4x\sqrt{-g}\Bigg[\frac{1}{16\pi}\Big\{\phi R-\frac{\omega(\phi)}{\phi}g^{ab}\nabla_a\phi \nabla_b\phi-V(\phi)\Big\}+L^{\rm(mat)}(\boldsymbol{g},\psi)\Big]~.
\label{SJ}
\end{eqnarray}
Here $L^{\rm(mat)}(\boldsymbol{g},\psi)$ corresponds to the Lagrangian of the external matter field dependent on the metric $g_{\mu \nu}$ and matter degree of freedom $\psi$, and the other part $L^{\rm (grav)}(\boldsymbol{g},\phi)$ is the gravitational Lagrangian consisting of the metric along with the kinetic and the potential terms of the scalar field $\phi$. Note that the coupling parameter $\omega(\phi)$ is in general a function of the scalar field $\phi$, when it is considered to be a constant, the above action corresponds to the Brans-Dicke theory of gravity. Moreover, in the term $\phi R$ in Eq.\ref{SJ}, the scalar field $\phi$ is non-minimally coupled with the Ricci scalar $R$. As a result, the gravitational interaction in this theory is mediated by the metric tensor $g_{ab}$ as well as the scalar field $\phi$. It is possible to transform the above action with non-minimal coupling to another minimal action, which requires the following transformation of the metric, 
\begin{align}
g_{ab}\rightarrow\widetilde{g}_{ab}=\phi g_{ab}~,
\label{GAB}
\end{align}
and, secondly, in order to reduce the kinetic term of the scalar field to a canonical form, one needs to re-scale the scalar field, such that,
\begin{align}
\phi\rightarrow\widetilde{\phi}\,\ {\textrm{with}}\,\ d\widetilde{\phi}=\sqrt{\frac{2\omega(\phi)+3}{16\pi}}\frac{d\phi}{\phi}~.
\label{PHI}
\end{align}
Using the above two transformation relations, the Jordan frame action in Eq.\ref{SJ}, transforms to the Einstein frame, where the non-minimal coupling is no longer present, and the action is given as
\begin{eqnarray}
&&\mathcal{A}_{\rm E}
=\widetilde{\A}^{\rm (grav)}[\boldsymbol{\t g},\t\phi]+\t{\A}^{\rm (mat)}[\boldsymbol{\t g},\t\phi,\psi]
=\int_{\mathcal{V}} d^4x\sqrt{-\widetilde{g}}\Big[\widetilde{L}^{\rm (grav)}(\boldsymbol{\t g},\t\phi)+\t{L}^{\rm (mat)}(\boldsymbol{\t g},\t\phi,\psi)\Big]
\no 
\\
&&\qquad=\int_{\mathcal{V}} d^4x\sqrt{-\widetilde{g}}\Big[\frac{\widetilde{R}}{16\pi}-\frac{1}{2}\widetilde{g}^{ab}\,\widetilde{\nabla}_a\widetilde{\phi}\,\widetilde{\nabla}_b\widetilde{\phi}-U(\widetilde{\phi})+\t{L}^{\rm (mat)}(\boldsymbol{\t g},\t\phi,\psi)\Big]~.
\label{SE}
\end{eqnarray}
The potential in the Einstein frame $U(\widetilde{\phi})$ and in the Jordan frame $V(\phi)$ are related by,  $U(\tilde{\phi})=\{V(\phi)/16\pi\phi^2\}$ and the matter Lagrangian in the two frames are connected as $\t{L}^{\rm (mat)}=\{L^{\rm (mat)}/\phi^2\}$. As already emphasized, Eq.\ref{SJ} is not the most general version of scalar-tensor theory. Instead, Horndeski and DHOST theories \cite{Horndeski:1974wa,Deffayet:2009wt,Kobayashi:2011nu} can be considered as the most general version of the scalar-tensor theory. However, due to the presence of higher derivative interactions involving $\phi$, non-minimal derivative coupling, and lack of Einstein frame representation, we have planned to attempt it separately in our future work. Further, note that the following analysis will also be valid for $f(R)$ theories of gravity \cite{Sotiriou:2008rp,DeFelice:2010aj}, which are a subclass of the scalar-tensor theories of gravity considered here. 

Ever since the formulation of this theory, there has been a major debate on whether these two frames, which are mathematically equivalent to each other, are also physically equivalent. If not, which of the frames is more physical. This debate has been going on for ages and has not yet been resolved (see \cite{Faraoni:1999hp, ALAL,Faraoni:1998qx, Faraoni:2010yi, Faraoni:2006fx, Saltas:2010ga, Capozziello:2010sc, Padilla:2012ze, Jacobson:1993pf, Deser:2006gt, Kamenshchik:2014waa, Banerjee:2016lco, Chakraborty:2016ydo, Chakraborty:2016gpg, Pandey:2016unk, Ruf:2017xon, Karam:2017zno, Bahamonde:2017kbs, Karam:2018squ, Chakraborty:2023kel, Bhattacharya:2023ycc}). The mathematical equivalence prompts one to relate thermodynamic quantities associated with one frame to another, which have been achieved recently for a certain set of thermodynamic parameters \cite{Bhattacharya:2017pqc, Bhattacharya:2018xlq, Bhattacharya:2022mnb} and here we wish to extend it further. It is well known that thermodynamics of gravity depends crucially on total derivative terms \cite{Wald:1993nt,Iyer:1994ys}, and the actions in the two frames, namely $\A_{\rm J}$ and $\A_{\rm E}$, differ by a boundary term, 
\begin{align}
\A_{\rm E}=\A_{\rm J}-\frac{3}{16\pi}\int_{\mathcal{V}}d^{4}x\sqrt{-g}\square\phi\equiv\A_{\rm J}' ~.
\label{ACTUAL}
\end{align}
The boundary term depending on $\square\phi$ does not affect the dynamics, however, since the thermodynamic parameters arise from the surface part of the action, they are not equivalent unless the contribution of the boundary term is taken into account \cite{Bhattacharya:2017pqc, Bhattacharya:2018xlq, Bhattacharya:2022mnb}. In the present case as well, we shall find that the $\square\phi$ term plays a crucial role in the equivalence of the two frames. To show this explicitly, we discuss three possible actions in the \ST theory of gravity --- (i) action of the Einstein frame $\A_{\rm E}$, (ii) action in the Jordan frame $\A_{\rm J}$, and (iii) action in the Jordan frame $\A'_{\rm J}$, which involves the total derivative term\footnote{One can argue that, why not modify the action of the Einstein frame by the $\square \phi$ term (as $\mg\square\phi=\tmg\t\square\ln\phi$), rather than modifying the Jordan frame action. Although, in doing so, the inequivalence at the level of action is removed, it raises issues concerning the decomposition of the action into bulk and surface parts, with appropriate thermodynamic properties (for more details, see \cite{Bhattacharya:2017pqc, Bhattacharya:2018xlq, Bhattacharya:2023ycc}). Hence we will not follow that route.}.

The dynamical equations for gravity and scalar can be determined in both of the frames by varying the actions with respect to $g_{ab}$ and $\phi$, respectively. This yields for the Jordan frame, 
\begin{align}
\delta \A_{\rm J}&=\int_{\mathcal{V}} d^4x\sqrt{-g}\left(E_{ab}\delta g^{ab}+E_{\phi}\delta\phi\right)
+\frac{1}{16\pi}\int_{\mathcal{V}}d^4x\,\partial_{i}\left[\phi\sqrt{-g}\left(g^{ab}\delta\Gamma^{i}_{ab}-g^{ij}\delta\Gamma^{k}_{jk}\right)\right]
\nonumber
\\
&\qquad +\frac{1}{16\pi}\int_{\mathcal{V}}d^4x\,\partial_{i}\left[\sqrt{-g}\partial_{j}\phi \delta g^{ij}
-\sqrt{-g}\partial^{i}\phi\,g_{ab}\delta g^{ab}\right]
-\frac{1}{8\pi}\int_{\mathcal{V}}d^4x\,\partial_{i}\left[\sqrt{-g}\frac{\omega(\phi)}{\phi}\partial^{i}\phi \delta \phi\right]~,
\end{align}
where, the field equations for gravity and field equations for scalar yields,
\begin{align}
E_{ab}&\equiv\phi G_{ab}+\frac{\omega}{2\phi}\nabla_i\phi\nabla^i\phi g_{ab}-\frac{\omega}{\phi}\nabla_a\phi\nabla_b\phi
+\frac{V}{2}g_{ab}-\nabla_a\nabla_b\phi+\nabla_i\nabla^i\phi g_{ab}-8\pi T^{\rm (mat)}_{ab}=0~,
\label{EoMST}
\\
E_{\phi}&\equiv \square\phi-\frac{1}{2\omega+3}\Bigg[8\pi T^{\rm (mat)}-\frac{d\omega}{d\phi}\nabla_i\phi\nabla^i\phi+\phi \frac{dV}{d\phi}-2V\Bigg]=0~.
\end{align}
Here $G_{ab}$ is the Einstein tensor of the Jordan frame, and the energy-momentum tensor $T_{ab}^{\rm (mat)}$ for the matter field is defined as
\begin{align}
T_{ab}^{\rm (mat)}=-\frac{2}{\mg}\frac{\p(\mg L^{\rm (mat)})}{g^{ab}}~.
\end{align}
Keeping in mind future application, we note the trace of the gravitational field equations, as presented in Eq.\ref{EoMST}, which read: $E\equiv g^{ab}E_{ab}=-\phi R+(\omega/\phi)\nabla_{i}\phi \nabla^{i}\phi+2V+3\square \phi-8\pi T^{\rm (mat)}$. Multiplying this expression by $(1/2)g_{ab}$ and subtracting from $E_{ab}$, we obtain the following alternative gravitational field equations, 
\begin{align}
\bar{E}_{ab}\equiv E_{ab}-\frac{1}{2}E g_{ab}&=\phi R_{ab}-\frac{\omega}{\phi}\nabla_a\phi\nabla_b\phi
-\frac{V}{2}g_{ab}-\nabla_a\nabla_b\phi-\frac{1}{2}\nabla_i\nabla^i\phi g_{ab}-8\pi \bar{T}^{\rm (mat)}_{ab}=0~,
\label{EoMSTalt}
\end{align}
where, $\bar{T}^{\rm (mat)}_{ab}=T_{ab}^{\rm (mat)}-(1/2)g_{ab}T^{\rm (mat)}$. We would like to emphasize that the same field equations for both gravity and the scalar sectors can be obtained from the variation of the action $\A_{\rm J}'$ as well. This can be seen from the following result, connecting the variation of the action $\A_{\rm J}'$ and the variation of the action $\A_{\rm J}$ as, 
\begin{align}
\delta \A_{\rm J}'=\delta \A_{\rm J}
-\frac{3}{16\pi}\int_{\mathcal{V}}d^{4}x\partial_{i}\left[\sqrt{-g}\partial_{j}\phi\left(\delta g^{ij}-\frac{1}{2}g^{ij}g_{ab}\delta g^{ab}\right)+\sqrt{-g}g^{ij}\partial_{j}\delta \phi\right]~.
\end{align}
As evident, the variations differ by total derivative terms and hence does not affect the dynamical equations. On the other hand, for Einstein frame, the variation of the action with respect to the metric and the scalar field yields, 
\begin{align}
\delta \A_{\rm E}&=\int_{\mathcal{V}} d^4x\sqrt{-\t g}\left(\t E_{ab}\delta \t g^{ab}+\t E_{\phi}\delta\t\phi\right)
+\frac{1}{16\pi}\int_{\mathcal{V}}d^4x\,\partial_{i}\left[\sqrt{-\t g}\left(\t g^{ab}\delta\t \Gamma^{i}_{ab}-\t g^{ij}\delta\t \Gamma^{k}_{jk}\right)\right]
\nonumber
\\
&\qquad -\int_{\mathcal{V}}d^4x\partial_{i}\left(\sqrt{-\t g}\,\t g^{ij}\partial_{j}\t \phi\,\delta\tilde\phi\right)~,
\end{align}
where, the dynamical equations for gravity and for scalar are provided as follows:
\begin{align}
\t E_{ab}&=\widetilde{G}_{ab}-8\pi\t T^{\rm (mat)}_{ab}
-8\pi\Big[\tilde{\nabla}_a\tilde{\phi}\tilde{\nabla}_b\tilde{\phi}-\frac{1}{2}\tilde{g}_{ab}\tilde{\nabla}^i\tilde{\phi}\tilde{\nabla}_i\tilde{\phi}-\tilde{g}_{ab}U(\tilde{\phi})\Big]=0~;
\label{eomgravein}
\\
\t E_{\phi}&=\t{\na}_a\t{\nabla}^a\tilde{\phi}-\frac{dU}{d\tilde{\phi}}=0~,
\label{eomscalarein}
\end{align}
with $\t G_{ab}$ being the Einstein tensor in the Einstein frame. Alike the Jordan frame, here also the trace of the gravitational field equations read: $E=-\t R-8\pi \t T^{\rm (mat)}+8\pi(\tilde{\nabla}^i\tilde{\phi}\tilde{\nabla}_i\tilde{\phi}+4U)$. Using which we can write down an alternative form for the gravitational field equations in the Einstein frame: 
\begin{align}\label{altgraveqein}
\widetilde{\bar{E}}_{ab}&\equiv E_{ab}-\frac{1}{2}g_{ab}E=\widetilde{R}_{ab}-8\pi\widetilde{\bar{T}}^{\rm (mat)}_{ab}-8\pi\Big[\tilde{\nabla}_a\tilde{\phi}\tilde{\nabla}_b\tilde{\phi}+\tilde{g}_{ab}U(\tilde{\phi})\Big]=0~,
\end{align}
where, $\widetilde{\bar{T}}^{\rm (mat)}_{ab}\equiv\t T^{\rm (mat)}_{ab}-(1/2)\t g_{ab}\t T^{\rm (mat)}$. Using the transformation properties of the Einstein tensor under Eq.\ref{GAB} and the transformation property of the scalar under Eq.\ref{PHI}, one can demonstrate the mathematical equivalence of the dynamical equations in the two frames of reference. In particular, we notice that the energy-momentum tensors in the two frames are related as,
\begin{equation}
\widetilde{T}^{\rm (mat)}_{ab}=\frac{1}{\phi}T^{\rm (mat)}_{ab}~,
\quad
\widetilde{T}^{a~\textrm{(mat)}}_{b}=\frac{1}{\phi^2}T^{a~\textrm{(mat)}}_{b}~,
\quad
\widetilde{T}^{ab~\textrm{(mat)}}=\frac{1}{\phi^3}T^{ab~\textrm{(mat)}}~.
\label{conformtab}
\end{equation}
These results involving the gravitational actions in different frames, associated boundary terms, the dynamical equations and the connection between the energy momentum tensors will be key for our subsequent analysis. We will now discuss the structure of the Noether current in these two frames of reference in the subsequent section.   

\subsection{Noether current in scalar-tensor theories of gravity}

The actions for the scalar-tensor theories of gravity, considered here, are described by Eq.\ref{SJ} in the Jordan frame and by Eq.\ref{SE} in the Einstein frame, are both diffeomorphism invariant. This guarantee the existence of a conserved Noether current under the transformation $x^{i}\to x^{i}+\xi^{i}(x)$, which plays a central role in black hole thermodynamics. For example, if the diffeomorphism vector field $\xi^{i}$ represents a Killing symmetry associated with time translation, the conserved charge, associated with the corresponding Noether current, evaluated at infinity is related to the mass. If the Killing symmetry corresponds to rotation, then the Noether charge at infinity connects with angular momentum, and most importantly, if the Killing symmetry is associated with the generators of a Killing horizon, then the Noether charge on the Killing horizon is associated with entropy of the horizon \cite{Wald:1993nt,Iyer:1994ys}. However, for both GR and Lovelock theories of gravity, even for a general diffeomorphism vector field, the conserved Noether charge has a thermodynamic interpretation when evaluated on a certain timelike hypersurface, as well as on a generic null surface, in terms of the surface heat content \cite{Padmanabhan:2013nxa,Chakraborty:2014rga,Chakraborty:2018qew,Chakraborty:2015aja}. In this work we wish to demonstrate that such an interpretation exists for the scalar-tensor theories of gravity as well, and in both the frames of reference. As a first step to these directions, we present below, the expressions of the (off-shell) Noether current due to diffeomorphism invariance of the scalar-tensor action following \cite{Bhattacharya:2018xlq}. 

\subsubsection{Noether current in the Jordan frame}

In the Jordan frame, we have two actions --- (a) the one described by Eq.\ref{SJ}, and (b) the modified action in Eq.\ref{ACTUAL}, differing by a total derivative. For the Jordan frame action in Eq.\ref{SJ}, the corresponding Noether current due to diffeomorphism invariance can be obtained as $J^a[v]=\na_bJ^{ab}[v]$, where the Noether potential $J^{ab}[v]$ reads (for more details, see \cite{Bhattacharya:2018xlq}),
\begin{align}
J^{ab}[v]=\frac{1}{16\pi}\Big[\phi(\nabla^av^b-\nabla^bv^a)+2v^a(\nabla^b\phi)-2v^b(\nabla^a\phi)\Big]~. 
\label{JABJORDAN}
\end{align}
On the contrary, if we consider the action $\mathcal{A}_{\rm J}'$ in the Jordan frame, the corresponding Noether current will differ by a total derivative term, such that the Noether potential becomes \cite{Bhattacharya:2018xlq},
\begin{align}
J'^{ab}[v]&=J^{ab}[v]+\frac{3}{16\pi}\left(-v^{i}\nabla^{j}\phi+v^{j}\nabla^{i}\phi\right)
\nonumber
\\
&=\frac{1}{16\pi}\left[\nabla^a(\phi v^b)-\nabla^b(\phi v^a)\right]~. 
\label{JAB'}
\end{align}
It is intriguing that the structure of the Noether potential $J'^{ab}$ associated with the Jordan frame action $\mathcal{A}_{\rm J}'$ is very similar to the Noether potential associated with the Einstein-Hilbert action, with the diffeomorphism vector field scaled by the scalar field $\phi$.

\subsubsection{Noether current in the Einstein frame}

In the Einstein frame as well, the Noether current $\t{J}^{a}$, arising from the diffeomorphism invariance of the action $\mathcal{A}_{\rm E}$, can be expressed in terms of the Noether potential $\t{J}^{ab}$, as $\t{J}^{a}=\widetilde{\nabla}_{b}\t{J}^{ab}$, where the Noether potential reads \cite{Wald:1993nt,Iyer:1994ys},
\begin{align}
\t{J}^{ab}[v]=\frac{1}{16\pi}\left[\widetilde{\nabla}^{a}v^{b}-\widetilde{\nabla}^{b}v^{a}\right]~. 
\label{JABEIN}
\end{align}
Using the conformal transformation of the metric: $\t{g}_{ab}=\phi g_{ab}$ and the connection, the above Noether potential in the Einstein frame can be related to the Noether potential in the Jordan frame, as $\t{J}^{ab}=(1/\phi^{2})J'^{ab}$. Note that the Noether potential associated with the diffeomorphism invariance of the Einstein frame action $\mathcal{A}_{\rm E}$ is connected to the Noether potential arising from the Jordan frame action $\mathcal{A}_{\rm J}'$, but differs from the Noether potential of $\mathcal{A}_{\rm J}$. 

Moreover, the expression for Noether currents in \ST theories of gravity have been obtained using the respective field equations in both the frames, and hence are on-shell. While, it is also possible to demonstrate that the same expressions can be obtained from purely geometric arguments without any reference to the field equations. This brings the above analysis for \ST theories of gravity at the same footing as that of GR and provides a completely geometrical origin for the Noether current, without invoking any symmetry argument (for the derivation, see \ref{appennoether}). In the following section, we show that for a generic diffeomorphism vector field, not necessarily a Killing vector, the conserved Noether charge has a thermodynamic interpretation in terms of the surface heat content. This can also be demonstrated to be related to the evolution of spacetime being due to the difference between appropriately defined surface and bulk degrees of freedom. 

\section{Noether charge as the surface heat content} \label{heatcontent}

Earlier in the context of general relativity, as well as Lanczos-Lovelock gravity, it has been found that the Noether charge within a given spacetime volume can be interpreted as the heat content of the surface associated with this volume \cite{Padmanabhan:2013nxa,Chakraborty:2014rga}. This result holds even for a general diffeomorphism vector field, which is not necessarily a Killing vector. In this section, we demonstrate that the above conclusion is valid for the scalar-tensor theories as well, involving non-minimal terms in the action, so that the Bekenstein-Hawking area law of entropy is not valid \cite{Kang:1996rj}. Yet, we find that the thermodynamic structure is still maintained, depicting the universality of gravity-thermodynamics correspondence. At the same time, keeping in mind the age-old debate regarding the physical equivalence of the Einstein and Jordan frames, we also discuss how the thermodynamic quantities of the two frames are related. 

Before jumping into the main analysis, let us summarize the spacetime foliation that we will be using in our work. We consider the spacetime to be foliated by a series of spacelike hypersurfaces which are defined by $t(x)=$constant. The unit normal corresponding to this hypersurface is becomes $u_a=-N\na_at$ (in Einstein frame $\t u_a=-\t N\t\na_at$), with $N$ (or $\t N$) being the lapse function. This will reduce to $u_a=-N\delta^0_a$ (or $\t u_a=-\t N\delta^0_a$) if one considers $t(x)$ as the temporal coordinate of the spacetime. Furthermore, the induced metric on the surface can be obtained as
\begin{align}
h_{ab}=g_{ab}+u_au_b ~~\textrm{(Jordan frame)},~~~~~~~~\t h_{ab}=\t g_{ab}+\t u_a\t u_b~~ \textrm{(Einstein frame)}.
\end{align}
The induced metrics as defined above, also play the role of projection tensor (as $u_ah^a_{b}=0$ and $h^a_bh^b_c=h^a_c$; which translates to Einstein frame as well), which projects every vector on the $t=\textrm{constant}$ hypersurface. Furthermore, this spacetime foliation also allows us to define another vector, which will act as the Killing vector for static spacetimes, and reads, 
\begin{align}
\xi^a=N u^a~\textrm{(Jordan frame)}~,\qquad \t\xi^a=\t N\t u^a~\textrm{(Einstein frame)}~. 
\label{diffjor}
\end{align}
As mentioned earlier, for a static spacetime, $\xi^a=(\partial/\partial t)^{a}$ will be the timelike Killing vector field. However, in our analysis, we do not consider the spacetime to be static, and maintain the generality. For simplicity, we start our analysis with the Einstein frame where the analysis is identical to that of the general relativity. Thereafter, we move on to the analysis in the Jordan frame, where non-triviality arises.

\subsection{Einstein frame} \label{subseceinfol}

The expression of the Noether current in the Einstein frame has been provided in Eq.~\ref{JABEIN}, which is identical to that of general relativity, as expected. Therefore, following \cite{Padmanabhan:2013nxa,Chakraborty:2014rga}, it can be shown that the total Noether charge contained in a region (within the $t=$constant  hypersurface), which is bounded by $\t N=$constant surface, can be obtained as \cite{Padmanabhan:2013nxa,Chakraborty:2014rga}
\begin{align}
\t Q=\int_{\R}d^3x\sqrt{\t h} \t u_a \t{J}^{a}[\t\xi]=\int_{\p\R} d^2x\frac{\sqrt{\t{\sigma}}}{4}\frac{\t N \t a}{2\pi}=\int_{\p\R}d^2x\,\t T_{\rm loc}\t s~, 
\label{noetherein}
\end{align}
where $\t a=\sqrt{\t a^i\t a_i}$ and $\t a^a$ is defined as $\t a^a=\t u^i\t \na_i \t u^a=\t h^{ai}\t\na_i \ln \t N=\t D^a\ln \t N$. Here, $\t D_{a}$ denotes covariant derivative on the $t=\textrm{constant}$ hypersurface. Furthermore, in Eq.\ref{noetherein}, $\t\sigma$ denotes the determinant of the induced metric $\t\sigma_{ab}\equiv \t g_{ab}+\t u_a\t u_b-\t r_a\t r_b$, with $\t r_{a}=\t a_{a}/\t a$ being the unit \emph{outward} normal to the $\t N=\textrm{constant}$ hypersurface, where $\t a$ is the magnitude of the acceleration four-vector\footnote{It is entirely possible that the \emph{outward} normal to the $N=\textrm{constant}$ surface is $-\t D_{a}\ln \t N$, in which case $\t r_{a}=-\t a_{a}/\t a$. This situation can be handled by inserting an overall factor of $\epsilon$ in Eq.\ref{noetherein}, where $\epsilon =\pm 1$, for the acceleration being directed outward/inward to the $N=\textrm{constant}$ hypersurface, respectively \cite{Chakraborty:2014rga}. Here for brevity, we have chosen $\epsilon=1$.}. In addition, $\t T_{\rm loc}\equiv (\t N\t a)/2\pi$ can be interpreted as the Tolman redshifted Unruh-Davies temperature for the observers with four-velocity $\t u_a=-\t N\delta^0_a$. Thus, in the Einstein frame, the total Noether charge contained in a volume bounded by the $\t N=\textrm{constant}$ hypersurface can be interpreted as the heat content of the surface which encloses the volume. We now move on to the Jordan frame to examine such possibility.

\subsection{Jordan frame} \label{subsecjorfol}

Earlier in the Einstein frame, we found that the Noether charge within the $\t N=\textrm{constant}$ hypersurface, corresponding to the diffeomorphism vector field $\t\xi^a$ can be interpreted as the surface heat content. Here, in the following, we check the same possibility, albeit in the Jordan frame of reference. Besides, we would also like to observe if the thermodynamic parameters have a one-to-one correspondence between the two frames. In particular, we will show that the Noether current in Eq.\ref{JAB'}, arising from the action in Eq.\ref{ACTUAL} has thermodynamic interpretation that smoothly carries over from the Einstein frame. While, the Noether current presented in Eq.\ref{JABJORDAN} does not have such a correspondence with the Einstein frame. This is due to the total derivative term in Eq.\ref{ACTUAL}, which leads to identical dynamics, but different thermodynamics. 

We first consider the Noether current corresponding to the modified action $\mathcal{A}_{\rm J}'$ in Eq.\ref{ACTUAL}, which is given in Eq.\ref{JAB'}. For the diffeomorphism vector field $\xi^{a}$, defined in Eq.\ref{diffjor}, the Noether current becomes,
\begin{align}
16\pi J'^a[\xi]=\na_b\Big[N\phi\Big(a^au^b-a^bu^a\Big)+u^b\na^a(N\phi)-u^a\na^b(N\phi)\Big]~.
\end{align}
The above expression is obtained by the use of the following results: $\na_au_b=-K_{ab}-u_aa_b$; $a^i\equiv u^{a}\nabla_{a}u^{i}=D^{i}\ln N$, where $K_{ab}$ is the extrinsic curvature and $D_{a}$ is the three dimensional covariant derivative associated with the $t=\textrm{constant}$ hypersurface. Further simplification of the Noether current yields the following expression for the combination $u_aJ'^a[\xi]$, which reads,
\begin{eqnarray}
16\pi u_aJ'^a[\xi]=D_a\Big[2N\phi a'^a\Big]~, 
\label{jaximid}
\end{eqnarray}
where, $a'^a$ is defined as,
\begin{align}
a'^a\equiv a^a+\frac{1}{2}D^a\ln\phi=D^a\ln(\sqrt{\phi}N)~. 
\label{a'}
\end{align}
The first part of the vector $a'^a$ as given in Eq.\ref{a'} can be identified as the acceleration of the observer moving normally to the $t=\textrm{constant}$ surface and the second term (containing $D^a\ln \phi$) is the extra contribution appearing due to the non-minimal coupling between scalar and gravity in the theory. It is intriguing that even in Jordan frame the Noether charge density can be expressed as a total derivative in three dimensions, and just like the acceleration, the vector $a'^{a}$ is expressible as a three dimensional gradient. 

The total Noether charge contained in a region (within $t=$const. hypersurface), corresponding to the Noether current $J'^a$, is defined as $Q'=\int_{\nu}\sqrt{h}u_aJ'^a[\xi] d^3x$. Since $u_aJ'^a[\xi]$ is a total derivative term, the Noether charge is obtained in terms of the surface integral for a given surface that encloses the region. Here we choose the same surface as of the Einstein frame, i.e., $\t N=\sqrt{\phi}N=$constant surface which encloses the region under consideration. The unit normal corresponding to this surface will be given as
\begin{align}
r_{a}=\frac{D_{a}(\sqrt{\phi}N)}{\sqrt{D_{i}(\sqrt{\phi}N)D^{i}(\sqrt{\phi}N)}}=\frac{D_{a}\ln(\sqrt{\phi}N)}{\sqrt{D_{i}\ln(\sqrt{\phi}N)D^{i}\ln(\sqrt{\phi}N)}}=\hat{a}'_{a}~,\label{ra}
\end{align}
which is the unit modified acceleration vector. As a result, one finds that $r_aa'^a=\sqrt{a'_aa'^a}=a'$, where $a'$ is the magnitude of the modified acceleration. Given the above acceleration, it follows that $T'_{\rm loc}=(Na'/2\pi)$ plays the role of the Tolman-redshifted Unruh-Davies temperature of the observers with four-velocity $u_a=-N\nabla_{a}t$. Unlike the Einstein frame, where the local (Unruh-Davies) temperature is related to the spacetime metric alone, in the Jordan frame we find that the additional contribution from $\phi$ appears due to the non-minimal coupling. This feature is also present in \cite{Faraoni:2022gry}, where the temperature in the Jordan frame is shown to depend explicitly on $\phi$.

Finally, integrating $u_aJ'^a[\xi]$ on the $t=$constant hypersurface, we obtain the total Noether charge bounded by the $\sqrt{\phi}N=$ constant hypersurface, to yield
\begin{align}
\int_{\R}\sqrt{h}u_aJ'^a[\xi] d^3x=\int_{\p\R}\frac{\phi\sqrt{\sigma}}{4}\frac{ N a'}{2\pi}d^2x=\int_{\p\R} T'_{\rm loc} sd^2x~, 
\end{align}
where the entropy density $s'=\phi\sqrt{\sigma}/4$ and the total entropy is simply the integration of $s'$ over the $N\textrm{constant}$ hypersurface. The equivalence between the two frames requires both the local temperature and the entropy density to be related with each other. The equality between the local temperatures can be seen as follows: under the conformal transformation, the lapse function, the normal vector and the diffeomorphism vector field transforms as, $\t N=\sqrt{\phi} N$, $\t u_a= \sqrt{\phi} u_a$, and $\t\xi^a=\xi^a$, respectively. Therefore, the accelerations in the two frames are related as
\begin{align}
\t a^a=\t h^{ab}\t\na_b(\ln\t N)=\frac{1}{\phi}D^{a}\ln (\sqrt{\phi}N)=\frac{a'^{a}}{\phi}~. 
\label{aatil}
\end{align}
As a consequence, it follows that the redshifted local temperatures in the two frames of reference are related, 
\begin{align}
\t T_{\rm loc}=\frac{\t N\t a}{2\pi}=\frac{N a'}{2\pi}=T'_{\rm loc}~.
\end{align}
Therefore, the temperature in both the frames are identical, which is consistent with previous literature \cite{Jacobson:1993pf}. Similarly, in the Einstein frame, we had defined the entropy density as $\t s=(\sqrt{\t\sigma}/4)$, which is related to the determinant of the conformal transformed two-metric as $\t s=\phi (\sqrt{\sigma}/4)=s'$ (note that $\t\sigma$ and $\sigma$ are related as $\t\sigma=\phi^{2}\sigma$). Hence we find that the entropy density and the total entropy are equivalent. 

To summarize, alike the case of Einstein gravity, in scalar-tensor theories of gravity as well, the Noether current corresponding to the modified action in Eq.\ref{ACTUAL}, yields a Noether charge that can be interpreted as the surface heat content. In addition, we find that both the entropy and the local temperature in the Jordan frame, associated with the action in Eq.\ref{ACTUAL}, are equivalent with the corresponding expressions in the Einstein frames. This brings up the natural question, what happens when we do not include the $\square\phi$ term in the analysis (as is usually done in literature), i.e., when we consider the Noether charge corresponding to the action $\mathcal{A}_{J}$ in Eq.\ref{SJ}? The answer is in affirmative, the Noether charge associated with the action $\mathcal{A}_{J}$ also has the thermodynamic interpretation in terms of the heat content of a suitable boundary surface. This has been presented in \ref{appenheatcontent}, however, in this case, we find that the thermodynamic parameters are not equivalent to the Einstein frame. This is expected as dynamical equivalence does not guarantee thermodynamic equivalence.  

\section{Gravitational dynamics and holographic equipartition} \label{equipartition}

In this section, we will connect the Noether current and its associated charge density with the dynamics of the spacetime for \ST theories of gravity. This process also allows us to naturally define an appropriate bulk and surface degrees of freedom, which are equal for static spacetimes. While, their difference characterizes the gravitational dynamics. Since the equality is between a bulk and a surface degree of freedom, it is referred to as the holographic equipartition. Again, for simplicity, we start with the analysis in the Einstein frame and then the computation in the Jordan frame follows.
\subsection{Einstein frame}

In the Einstein frame, the analysis will be similar to \gr\,(which can be found in \cite{Padmanabhan:2013nxa,Chakraborty:2014rga}). We mention the key results, which will be necessary for comparison with the corresponding expressions in the Jordan frame. The evolution equation, describing gravitational dynamics in terms of the difference between bulk and surface degrees of freedom is given by \cite{Padmanabhan:2013nxa,Chakraborty:2014rga}, 
\begin{align}
\frac{1}{8\pi}\int_{\R}d^3x\,\t h_{ab} \tlie\t P^{ab}=\frac{1}{2}\k_{\rm B} \t T_{\rm avg}\Big(\t n_{\rm bulk}-\t n_{\rm sur}\Big)~. 
\label{holrelein23}
\end{align}
Here, $\t P^{ab}$ denotes the momentum conjugate to the dynamical variable $\t h_{ab}$\footnote{As is well known, in Einstein gravity, the spatial metric is the dynamical one, the lapse and the shift functions are non-dynamical \cite{Arnowitt:1960es, Chakraborty:2016yna}. Hence we consider the momentum conjugate to the dynamical part of the metric, namely $h_{ab}$ alone.}, which is given as \cite{Arnowitt:1960es}
\begin{align}
16\pi \t P^{ab}=\sqrt{\t h}\Big[\t h^{ab}\t K-\t K^{ab}\Big]~. 
\label{PABTILEIN}
\end{align}
Besides the momentum, the above expression involves three additional quantities, the bulk degrees of freedom $\t n_{\rm bulk}$, the surface degrees of freedom $\t n_{\rm sur}$ and the redshifted temperature $T_{\rm loc}$ averaged over the $\t N=\textrm{constant}$ hypersurface, defined as
\begin{align}
\t T_{\rm avg}=\frac{1}{\t A}\int_{\partial \R}d^2x\sqrt{\t\sigma} \t T_{\rm loc}~,
\quad
\t n_{\rm sur}=\t A=\int_{\p\R}d^2x\sqrt{\t\sigma}~,
\quad
\t n_{\rm bulk}=\frac{1}{\frac{1}{2}\k_B\t T_{\rm avg}}\int_{\R}d^3x\sqrt{\t h}\t\rho_{\textrm{Komar}}~,
\end{align}
where the Komar energy density has the following expression:
\begin{align}
\t \rho_{\rm Komar}=\frac{\t N}{4\pi} \t R_{ab}\t u^{a}\t u^{b}=2 \t N \widetilde{\bar{T}}_{ab}^{\rm (mat)}\t u^{a}\t u^{b}+2 \t N\left(\t u^{a}\t \nabla_{a}\t \phi\right)^{2}-2 U(\t \phi)~,
\end{align}
where, we have used Eq.\ref{altgraveqein}. Besides, we have also used the following definition:  $\widetilde{\bar{T}}^{\rm (mat)}_{ab}=\t T^{\rm (mat)}_{ab}-(1/2)\t g_{ab}\t T^{\rm (mat)}$, where $\t T^{\rm (mat)}_{ab}$ is the matter energy-momentum tensor and $\t T^{\rm (mat)}$ is its trace in the Einstein frame. Note that if we use the relations:
\begin{align}
\widetilde{\bar{T}}_{ab}^{\rm (mat)}&=\frac{1}{\phi}\bar{T}_{ab}^{\rm (mat)}~,
\quad
\t u^{a}=\frac{1}{\sqrt{\phi}}u^{a}~,
\quad
U(\t \phi)=\frac{V(\phi)}{16\pi \phi^{2}}~,
\nonumber
\\
\t N&=\sqrt{\phi}N~,
\quad
\t u^{a}\t \nabla_{a}\t \phi=\frac{1}{\phi\sqrt{\phi}}\left(u^{a}\nabla_{a}\phi\right)\sqrt{\frac{2\omega(\phi)+3}{16\pi}}~,
\end{align}
the Komar energy density reads,
\begin{align}\label{KomarEinalt}
\t \rho_{\rm Komar}=\frac{2N}{\phi\sqrt{\phi}}\bar{T}_{ab}^{\rm (mat)}u^{a}u^{b}+\frac{2\omega(\phi)+3}{8\pi \phi^{2}\sqrt{\phi}}N\left(u^{a}\nabla_{a}\phi\right)^{2}-\frac{V(\phi)}{8\pi \phi^{2}}~,
\end{align}
which expresses the Komar energy density in the Einstein frame, in terms of quantities in the Jordan frame, which we will use in the subsequent sections to show the equivalence between the two. 

The interpretation of Eq.\ref{holrelein23} goes as follows: the Lie-variation of the canonical momentum conjugate to the dynamical variable in Einstein gravity, is related to the difference between the above defined bulk and surface degrees of freedom. The surface degrees of freedom needs no explanation, as the gravitational degrees of freedom adds up to the area of the horizon. The bulk degrees of freedom, on the other hand, is obtained from the equipartition of the Komar energy density within the $N=\textrm{constant}$ hypersurface\footnote{A few comments are in order. Even though the number of surface degrees of freedom $n_{\rm sur}$ is manifestly positive, the Komar energy density is not necessarily positive, and hence to make the number of bulk degrees of freedom positive, one may introduce a factor of $\epsilon$ in the definition of $n_{\rm bulk}$. Note that negative Komar energy density is associated with the violation of energy condition, which changes the attractive nature of gravity and hence the sign of $D_{a}N$. Thus the directions of $r_{a}$ and $a_{a}$ are opposite, only when the Komar energy density becomes negative. Hence there will be an overall factor of $\epsilon$ in the right hand side of Eq.\ref{holrelein23}, which again we omit for brevity.}. 

When the spacetime is static, the diffeomorphism vector field $\t\xi^a$ reduces to a timelike Killing vector field and hence the Lie derivative term $\tlie\t P^{ab}$ vanishes. As a result, one obtains the holographic equipartition relation, 
\begin{align}
\t n_{\rm bulk}=\t n_{\rm sur}~,
\label{holeqein}
\end{align}
implying the number of bulk and surface degrees of freedom coincides. For dynamical spacetime, there is a deviation from the equipartition relation, and the difference between surface and bulk degrees of freedom drives the evolution of the spacetime, through non-zero Lie derivative of the gravitational momentum. We now move on to the Jordan frame, where a scalar field is non-minimally coupled to gravity, and wish to find out the status of the equipartition as well as evolution of the spacetime geometry. 

\subsection{Jordan frame}

In the Jordan frame, we start with the basic identity that has been central to the derivation of equipartition and spacetime evolution in \gr. This follows from the following geometrical identity (for a derivation of this identity, see Eq.\ref{danaajor} of \ref{appennoether}):
\begin{align}
2\sqrt{h} D_a\Big[ N a^a\Big]=2\sqrt{h}N R_{ab} u^a u^b-16\pi  h_{ab} \lie P^{ab}~.
\label{idenjor}
\end{align}
Note that the term $R_{ab}u^{a}u^{b}$, when field equations for Einstein gravity are used, reduces to $\bar{T}_{ab}u^{a}u^{b}$, corresponding to the Komar energy density. The tensor $P^{ab}$ in the above expression is defined as $16\pi P^{ab}=\sqrt{h}(h^{ab}K-K^{ab})$, which is related to the momentum in the scalar-tensor theory of gravity, denoted by $P^{ab}_{\rm (st)}$ \cite{Bhattacharya:2023ycc}, conjugate to the spatial metric $h_{ab}$ as,
\begin{align}
16\pi P^{ab}_{\rm (st)}=\sqrt{h}\Big[\Phi\Big(h^{ab} K-K^{ab}\Big)-h^{ab}u^i\na_i\phi\Big]=16\pi\phi P^{ab}-\sqrt{h}h^{ab}u^i\na_i\phi~.
\label{PABST}
\end{align}
Given this relation, it is a short exercise to establish the connection between the Lie-variation of these two quantities, yielding,
\begin{align}
16\pi h_{ab} \lie P^{ab}_{\rm (st)}=16\pi\phi  h_{ab} \lie P^{ab}+3\sqrt{h}K\lie\phi-3\sqrt{h}\lie\Big(u^i\na_i\phi\Big)~.
\label{liemomst}
\end{align}
Here, we have used the result that the Lie derivative of the induced metric $h_{ab}$ is directly proportional to the extrinsic curvature $K_{ab}$. Motivated by the structure of the above expression, we multiply Eq.\ref{idenjor} by a factor of $\phi$ and rewrite the same by substituting for $\phi  h_{ab} \lie P^{ab}$ from Eq.\ref{liemomst}, from which we obtain,
\begin{align}
16\pi h_{ab} \lie P^{ab}_{\rm (st)}&=2\sqrt{ h} N\phi R_{ab} u^a u^b
+3\sqrt{h}K\lie\phi-3\sqrt{h}\lie\Big(u^i\na_i\phi\Big)
\nonumber
\\
&\qquad +2\sqrt{h}N a^aD_a\phi-2\sqrt{h} D_a\Big[\phi N a^a\Big]~. 
\label{identityjor}
\end{align}
In the previous section, while discussing the thermodynamic interpretation of the Noether charge in the Jordan frame, we noticed that it is not the acceleration itself, rather a modified acceleration $a'_{a}=D_{a}\ln (\sqrt{\phi}N)$, which plays the key role. In addition, we have found that $u_aJ'^a[\xi]$ is a total three-derivative, with the following form $D_a[N\phi a'^a]$. Therefore, instead of $a^{a}$, we wish to express Eq.\ref{identityjor} in terms of $a'^{a}$. This requires determining the connection between $D_a[N\phi a'^a]$ and $D_a[\phi N a^a]$, which becomes, (see \ref{appenconnection} for a derivation)
\begin{align}
D_a\Big[\phi N a^a\Big]=D_a\Big[N\phi a'^a\Big]-\frac{N}{2}\square\phi+\frac{K}{2}\lie\phi-\frac{1}{2}\lie\Big(u^i\na_i\phi\Big)~.
\label{conenction}
\end{align}
Substituting Eq.\ref{conenction} in Eq.\ref{identityjor}, thereby replacing $a^a$ by $a'^a$ inside spatial derivative, we obtain the following relation,
\begin{align}\label{intmrelation1}
16\pi h_{ab} \lie P^{ab}_{\rm (st)}&=2\sqrt{h} N\phi R_{ab} u^a u^b +N \sqrt{h}\square \phi
+2\sqrt{h}K\lie\phi-2\sqrt{h}\lie\Big(u^i\na_i\phi\Big)
\nonumber
\\
&\qquad +2\sqrt{h}N a^aD_a\phi-2\sqrt{h}D_{a}\Big[\phi N a'^{a}\Big]~. 
\end{align}
Using the result, $\lie(u^{i}\na_{i}\phi)=\xi^{j}\nabla_{j}(u^{i}\na_{i}\phi)=Nu^{a}u^{b}\nabla_{a}\nabla_{b}\phi+Na^{i}D_{i}\phi$, and rearranging various terms of Eq.\ref{intmrelation1}, we arrive at the following expression, 
\begin{align}\label{intmrelation2}
16\pi h_{ab} \lie P^{ab}_{\rm (st)}&+\sqrt{h}\left(3u^{a}\nabla_{a}\ln \phi-2K\right)\lie\phi
=2N\sqrt{h}\left(\phi R_{ab}-\frac{1}{2}g_{ab}\square\phi-\nabla_{a}\nabla_{b}\phi+\frac{3}{2\phi}\nabla_{a}\phi\nabla_{b}\phi \right)u^au^b 
\nonumber
\\
&\qquad \qquad \qquad -2\sqrt{h}D_{a}\Big[\phi N a'^{a}\Big]~. 
\end{align}
Note that the left hand side of the above expression involves evolution of the gravitational degrees of freedom through the Lie variation of the momentum conjugate to the spatial metric $h_{ab}$, similarly, it also involves evolution of the scalar degree of freedom through the $\lie \phi$ term. This is expected, as in the scalar tensor theory, in particular in the Jordan frame, where gravity is non-minimally coupled with scalar, dynamical evolution of both the metric and the scalar field must be taken into account. Defining,
\begin{align}
P_{(\phi)}\equiv\frac{1}{16\pi}\left(3u^a\na_a\ln\phi-2K\right)~,
\end{align}
we can rewrite Eq.\ref{intmrelation2} as, 
\begin{align}\label{intmrelation3}
h_{ab} \lie P^{ab}_{\rm (st)}&+\sqrt{h}P_{(\phi)}\lie\phi
=\frac{N\sqrt{h}}{8\pi}\left(\phi R_{ab}-\frac{1}{2}g_{ab}\square\phi-\nabla_{a}\nabla_{b}\phi+\frac{3}{2\phi}\nabla_{a}\phi\nabla_{b}\phi \right)u^au^b-\frac{\sqrt{h}}{8\pi}D_{a}\Big[\phi N a'^{a}\Big]~. 
\end{align}
This is the dynamical evolution equation for scalar-tensor theory in the Jordan frame that we were after. The left hand side provides the dynamical evolution of the gravity and scalar degrees of freedom, while the first term on the right hand side depicts the Komar energy density in the Jordan frame, and the second term is related to the entropy density and local temperature. Integrating the above equation over $t=\textrm{constant}$ hypersurface, we obtain, 
\begin{align}\label{intmrelation4}
\int_{\mathcal{R}} d^{3}x~\Big(h_{ab} \lie P^{ab}_{\rm (st)}&+\sqrt{h}P_{(\phi)}\lie\phi\Big)
=\frac{1}{2}\int_{\mathcal{R}}d^{3}x\,\sqrt{h}\rho_{\rm Komar}'-\frac{1}{8\pi}\int_{\partial \mathcal{R}}d^{2}x\,\sqrt{\sigma}\phi N r_{a}a'^{a}~,
\end{align}
where, $\partial \mathcal{R}$ is the $N\sqrt{\phi}=\textrm{constant}$ surface within the $t=\textrm{constant}$ hypersurface and we have introduced the following Komar energy density,
\begin{align}\label{KomarJordan}
\rho_{\rm Komar}'=\frac{N}{4\pi}\left(\phi R_{ab}-\frac{1}{2}g_{ab}\square\phi-\nabla_{a}\nabla_{b}\phi+\frac{3}{2\phi}\nabla_{a}\phi\nabla_{b}\phi \right)u^au^b~.
\end{align}
Note that the above Komar energy density, besides having contribution from gravity also depends explicitly on scalar degrees of freedom. The vector $r_{a}$ is the unit outward normal to $N\sqrt{\phi}=\textrm{constant}$ surface, and as we have described before (see the discussion around Eq.\ref{ra} in \ref{subsecjorfol}), it is along the direction of the modified acceleration $a'^{a}$. Therefore, $r_{a}a'^{a}=|a'|$, and hence Eq.\ref{intmrelation4} reduces to, 
\begin{align}\label{intmrelation5}
\int_{\mathcal{R}} d^{3}x~\Big(h_{ab} \lie P^{ab}_{\rm (st)}&+\sqrt{h}P_{(\phi)}\lie\phi\Big)
=\frac{1}{2}\int_{\mathcal{R}}d^{3}x\,\sqrt{h}\rho_{\rm Komar}'-\int_{\partial \mathcal{R}}d^{2}x\,\frac{\phi\sqrt{\sigma}}{4}\frac{Na'}{2\pi}~.
\end{align}
Following \ref{subsecjorfol}, the local temperature associated with the observer having four-velocity $u_{a}$ is given by $k_{\rm B}T'_{\rm loc}=(Na'/2\pi)$ and the entropy density of scalar-tensor theory is $s'=\phi\sqrt{\sigma}/4$. This suggests the following definitions for the average temperature, surface degrees of freedom and the bulk degrees of freedom as, 
\begin{align}\label{tavgnsurnbulk}
T_{\rm avg}'\equiv \frac{\int_{\p\R}\sqrt{\sigma}\phi T'_{\rm loc}d^2x}{\int_{\p\R}\sqrt{\sigma}\phi d^2x}~;
\quad
n_{\rm sur}'=\int_{\partial \mathcal{R}}d^{2}x\,\phi\sqrt{\sigma}~;
\quad
n_{\rm bulk}'=\frac{2}{k_{\rm B}T_{\rm avg}'}\int_{\mathcal{R}}d^{3}x\,\sqrt{h}\rho_{\rm Komar}'~.
\end{align}
Note that, the temperature is being averaged over the two-surface $N\sqrt{\phi}=\textrm{constant}$, and also the measure is the entropy density, hence involves factors of $\phi$. The number of surface degrees of freedom is simply the area of the two-surface $N\sqrt{\phi}=\textrm{constant}$, and the bulk degrees of freedom are defined with respect to the Komar energy density within the two-surface. Thus we finally obtain, 
\begin{align}
\int_{\R}d^{3}x\Big[h_{ab}\lie P^{ab}_{\rm (st)}+\sqrt{h}P^{(\phi)}\lie\phi\Big]=\frac{\k_{\rm B}T'_{\rm avg}}{4}\Big(n'_{\rm bulk}-n'_{\rm sur}\Big)~. 
\label{holographicequipartition}
\end{align}
Thus the difference between the bulk and the surface degrees of freedom is responsible for the evolution of spacetime as well as of the scalar field, in the Jordan frame of scalar-tensor theory as well. Similarly, for static spacetimes, all the Lie variations appearing in the left hand side of the above equation vanishes, implying, $n'_{\rm bulk}=n'_{\rm sur}$. 

To see the equivalence of the above relation in the Jordan frame with the corresponding one in the Einstein frame, we start with the transformation of $R_{ab}$ under conformal transformation, which reads,
\begin{align}
\phi R_{ab}-\frac{1}{2}g_{ab}\square\phi=\phi \t R_{ab}-\frac{3}{2\phi}(\na_a\phi)(\na_b\phi)+\na_a\na_b\phi~.
\label{conform}
\end{align}
Using this relation, the Komar energy density in the Jordan frame, defined in Eq.\ref{KomarJordan}, reduces to, 
\begin{align}
\rho_{\rm Komar}'=\frac{N}{4\pi}\phi \t R_{ab} u^{a}u^{b}=\phi\sqrt{\phi}\t \rho_{\rm Komar}~.
\end{align}
Another way to arrive at this relation is through the use of the gravitational field equations in the Jordan frame, presented in Eq.\ref{EoMSTalt}, which reduces the Komar energy density to, 
\begin{align}
\rho_{\rm Komar}'=\frac{N}{4\pi}\left(8\pi \bar{T}^{\rm (mat)}_{ab}+\frac{2\omega+3}{2\phi}\nabla_a\phi\nabla_b\phi
+\frac{V}{2}g_{ab} \right)u^{a}u^{b}~.
\end{align}
As evident from Eq.\ref{KomarEinalt}, the above expression also yields the same relation between Komar energy densities in the two frames of reference, i.e., $\rho_{\rm Komar}'=\phi\sqrt{\phi}\t \rho_{\rm Komar}$. Thus Komar energy density, among the two frames of reference are connected by an overall factor involving the scalar field. Using the result $\t g_{ab}=\phi g_{ab}$, it follows that $\sqrt{\t h}=\phi\sqrt{\phi}\sqrt{h}$, and hence it follows that, $\sqrt{\t h}\t \rho_{\rm Komar}=\sqrt{h}\rho_{\rm Komar}'$.  

Following the previous discussion involving the connection between Noether charges in the two frames of reference, it is clear that the local temperatures are equivalent, i.e., $T'_{\rm loc}=\t T_{\rm loc}$. Further, it also follows that $s'=s$. Hence, from the definition of the average temperature in Eq.\ref{tavgnsurnbulk}, it follows that $T'_{\rm avg}=\t T_{\rm avg}$. Moreover, since $s=s'$, it follows that the surface degrees of freedom are the same in the two frames of reference: $n_{\rm sur}'=\t n_{\rm sur}$. Similarly, using the result that the Komar energy density times the integration measure are the same in the two frames of reference, as well as the equivalence of the average temperature, we obtain, $n'_{\rm bulk}=\t n_{\rm bulk}$.  

Thus we have not only depicted the validity of the result that departure from the holographic equipartition (\textit{i.e.} $n'_{\rm bulk}\neq n'_{\rm sur}$) leads to the time evolution of the spacetime as well as the scalar degrees of freedom in the Jordan frame, but also showed that the thermodynamic parameters (like entropy, local temperature, average temperature) as well as the bulk and the surface degrees of freedom are equivalent in the two frames. 

The above analysis has been performed with the Noether current $J'^a$, which corresponds to the modified action $\mathcal{A}_{\rm J}'$ in Eq.\ref{ACTUAL}. However, if we consider $J^a$ instead of $J'^a$, we will still obtain similar thermodynamic relations, but the thermodynamic parameters, as well as the bulk and the surface degrees of freedom will not be equivalent to that of the Einstein frame. The analysis of this inequivalent scenario has been provided in the \ref{appenholeq}, and the reason for the inequivalent description is because the actions $\mathcal{A}_{\rm J}$ and $\mathcal{A}_{\rm E}$ differ by total derivatives. Even though there is no difference dynamically, but at the thermodynamic level they can be distinguished. 

Thus, in the Jordan frame, one can obtain the holographic equipartition relation (and, thereby, connect it to the time-evolution of the spacetime) via two different routes, using different Noether currents $J^a[\xi]$ and $J'^a[\xi]$. In both the cases, it can be shown that the time-evolution in spacetime arises due to a difference between suitably defined surface and bulk degrees of freedom. However, in one case, we find that the physical parameters are equivalent, to the Einstein frame, while in the other, some of the thermodynamic parameters are not equivalent to the Einstein frame. Therefore, if one demands that not just dynamics, but also thermodynamical description be equivalent under the conformal transformation, one has to consider $\mathcal{A}_{\rm J}'$ as the more appropriate action in the Jordan frame than $\mathcal{A}_{\rm J}$.

\section{Conclusions} \label{concl}

The major goal of the present work is to draw the connection between gravitational dynamics and spacetime thermodynamics using the Noether charge formalism for scalar-tensor theories of gravity. It is well-known that gravitational dynamics of \gr, as well as Lovelock class of theories can be given a thermodynamic interpretation. In particular, the evolution of spacetime in these theories can be related to the difference between suitably defined bulk and surface degrees of freedom, where the bulk degrees of freedom are related to the Komar energy density and the surface degrees of freedom are related to the Wald entropy. In this work we demonstrate that such a thermodynamical description for gravitational dynamics exists for scalar-tensor theories of gravity as well. Moreover, we have also demonstrated that the Noether charge in the scalar-tensor theory of gravity indeed has the interpretation as the surface heat content, as was the case for \gr\ and Lovelock theories of gravity. Subsequently, given the fact that the scalar-tensor theories of gravity can be described either in the Einstein frame, or, in the Jordan frame, we have  explored if the thermodynamic description holds true in both the frames, and if these are equivalent.  

Our analysis and its implications, in view of the above discussion, can be summarized as follows: We first demonstrate that there are two possible actions in the Jordan frame, namely $\mathcal{A}_{\rm J}$ and $\mathcal{A}_{\rm J}'$, differing by a total derivative term. Among these two, the action $\mathcal{A}_{\rm J}'$ transforms to the Einstein frame action $\mathcal{A}_{\rm E}$ under conformal transformation, while $\mathcal{A}_{\rm J}$ differs from $\mathcal{A}_{\rm E}$ by a total derivative. Thus the variation of the actions $\mathcal{A}_{\rm J}$ and $\mathcal{A}_{\rm J}'$ yield the same dynamical equations for gravity and scalar, which transforms to the field equations derived from the variation of the Einstein frame action $\mathcal{A}_{\rm E}$ through conformal transformation. Hence as far as gravitational dynamics is considered, all of these actions are equivalent. 

Subsequently, starting from the action, we derived the conserved Noether current in both of these frames. Since Noether current is sensitive to the addition of a boundary term to the action, it follows that in the Jordan frame, the Noether currents associated with the actions $\mathcal{A}_{\rm J}$ and $\mathcal{A}_{\rm J}'$ are different, and only the Noether current obtained from $\mathcal{A}_{\rm J}'$ is equivalent with the Noether current in the Einstein frame. Using these expressions for the Noether currents, we find that the total Noether charges contained in a spatial volume, enclosed by a judiciously chosen surface, can be interpreted as the heat content of the surface. This interpretation is valid for both the frames and for both the actions $\mathcal{A}_{\rm J}$ and $\mathcal{A}_{\rm J}'$ in the Jordan frame. The entropy density in the Jordan frame depends on the scalar field and is given by $s=(\phi\sqrt{\sigma}/4)$, where $\sigma$ is the determinant of the boundary metric enclosing the spatial volume. The local Unruh temperature, on the other hand, is given by the magnitude of $D_{a}\ln (\sqrt{\phi}N)$, for $\mathcal{A}_{\rm J}'$, and is the magnitude of $D_{a}\ln (N/\phi)$ for $\mathcal{A}_{\rm J}$. Thus even though both of them gave identical thermodynamic interpretation, some of the thermodynamic parameters are different, e.g., temperature. Among these two distinct thermodynamic descriptions, the one originating from $\mathcal{A}_{\rm J}'$ maps to the results in the Einstein frame. Hence gravitational dynamics is not able to distinguish between Jordan and Einstein frame actions, but thermodynamics can do that. The thermodynamics arising out of $\mathcal{A}_{\rm J}$ and $\mathcal{A}_{\rm E}$ are distinct. 

Proceeding further, we demonstrate that the gravitational dynamics in each frames can be connected to the difference between naturally defined bulk and surface degrees of freedom. The static spacetimes, in particular, corresponds to the situation where the bulk and the surface degrees of freedom are equal, i.e., reaching the equipartition. Any departure from this relation leads to time evolution in the spacetime, as well as of the scalar field. Again, in the Jordan frame, one can find that both $\mathcal{A}_{\rm J}$ and $\mathcal{A}_{\rm J}'$ lead to an identical interpretation with different surface and bulk degrees of freedom. Among them the surface and the bulk degrees of freedom associated with the action $\mathcal{A}_{\rm J}'$ neatly maps into the corresponding degrees of freedom in the Einstein frame. While the degrees of freedom connected with $\mathcal{A}_{\rm J}$ are different from those of the Einstein frame. Thus the thermodynamic interpretation of gravitational dynamics holds true for scalar tensor theories, irrespective of the choice of the frames. However, the thermodynamic descriptions in the Jordan frame are not necessarily equivalent with the description in the Einstein frame.

Finally, we note that in the \ST theory of gravity, the extra scalar degrees of freedom, that couples non-minimally with the metric, modifies the dynamics and thermodynamics in a non-trivial manner. Even though the non-minimal coupling is simply $\phi R$, it neither appears in dynamics, nor in thermodynamics as an overall factor, rather affects all the expressions in a highly non-linear manner (see, e.g., the expressions of the field equations, Noether current, conjugate momenta to the spatial metric in the Jordan frame). Moreover, unlike the case of Einstein gravity, the equipartition relation involves not only the evolution of the spatial metric, but also the evolution of the scalar field. Furthermore, this work explicitly demonstrates that, while a total derivative term in the action does not affect the dynamics, it affects the thermodynamical quantities extensively. The above results are also applicable in the context of $f(R)$ theories of gravity, with $\phi$ replaced by $f'(R)$. Thus all of these results, originally derived in the context of general relativity, holds in a much broader context, from Lovelock gravity to scalar-tensor theories, including $f(R)$ theories of gravity. This depicts that the thermodynamic route to gravitational field equations is much deeper rooted than previously thought. 

There are a few future directions of exploration. In the present work we have mainly discussed about the non-minimal coupling between scalar and gravity through the simplest $\phi R$ term, and have kept the scalar part of the action quadratic in the first derivative. However, in general, there can be more complex couplings between gravity and scalar, as well as there can be higher derivatives of the scalar field present in the action, e.g., the Horndeski theories. Certain sub-classes of these theories also involve BH solutions with scalar hair, and hence it would be interesting to study their thermodynamic properties following the path laid out here. Moreover, the Horndeski-type Lagrangians are not specific to the scalar gravity coupling, as there can be similar Lagrangians involving non-minimal coupling and higher derivative terms involving vector fields as well. It would be interesting to see whether such theories can also have a thermodynamic makeover. We leave these issues for the future.  

\section*{Acknowledgement}

Part of the work of KB has been completed during the tenure of JSPS postdoctoral position in Fukushima University (JSPS KAKENHI Grant Number: 23KF0008). Research of SC is supported by MATRICS (MTR/2023/000049) and Core Research (CRG/2023/000934) Grants from SERB, Government of India. 

\appendix
\labelformat{section}{Appendix #1} 
\labelformat{subsection}{Appendix #1}
\section{Noether current from a geometric perspective}\label{appennoether}

In this appendix, we demonstrate that the Noether current can be derived from a purely geometrical point of view, without refereeing to any symmetry arguments whatsoever for both Einstein and the Jordan frames. 

The geometric derivation of the Noether current in the Einstein frame goes as follows: we first define the symmetric and the anti-symmetric parts of the combination $\tna^a \t v^b$, 
\begin{align}
\t S^{ab}[\t v]=\t \na^a\t v^b+\t \na^b\t v^a
\qquad
\qquad
\textrm{and}
\qquad
\qquad
\t A^{ab}[\t v]=\t \na^a\t v^b-\t \na^b\t v^a~. 
\label{symasymein}
\end{align}
Using properties of the covariant derivatives it is strightforward to show that $\t \na_{a}\t \na_{b}\t A^{ab}=0$, and hence the Noether current reads $16\pi\t J^{a}[\t v]\equiv \na_{b}\t A^{ab}$. The above also provides the Noether potential in the Einstein frame, presented in Eq.\ref{JABEIN}. This arises from purely geometrical standpoint without invoking any action. Its explicit form can also be determined in a similar fashion. From the above definitions, one obtains
\begin{align}
\t \na^a\t v^b=\frac{1}{2}\Big(\t S^{ab}[\t v]+\t A^{ab}[\t v]\Big)
\qquad
\qquad
\textrm{and}
\qquad
\qquad
\t\na_i\t v^i=\frac{1}{2}\t S[\t v]~.
\label{revsymasymein}
\end{align}
Using the above results to replace $\t\na^a\t v^b$ and $\t\na_b\t v^b$ by appropriate quantities, defined above, in the standard identity: $\t\na_b\t\na^a\t v^b-\t\na^a\t\na_b\t v^b=\t R^a_b\t v^b$, we obtain, 
\begin{align}
\t\na_b\t A^{ab}[\t v]+\t\na_b\Big[\t S^{ab}[\t v]-\t g^{ab}\t S[\t v]\Big]=2\t R^a_b\t v^b~.
\end{align}
Using the result that $\t\na_b\t A^{ab}[\t v]=16\pi \t J^a[\t v]$, and also the fact that $\t\na_b\t S^{ab}[\t v]$ is related to the Lie-variation of the metric tensor as: $\t S^{ab}[\t v]=-\pounds_{\t v}\t g^{ab}$, one finally obtains
\begin{align}
\t J^a[\t v]=\frac{1}{8\pi}\t R^a_b\t v^b+\frac{1}{16\pi}\t\na_b\Big[\pounds_{\t v}\t g^{ab}-\t g^{ab}\t g_{ij}\pounds_{\t v}\t g^{ij}\Big] 
\label{NCasGEOMEIN}
\end{align}
Using $\t v^{a}=\t \xi^{a}=\t N \t u^{a}$, and contracting the above expression with $\t u^{a}$, we obtain, 
\begin{align}
\t D_a(\t N \t a^a)=\t N\Big[\t R_{ab}\t u^a \t u^b+\t K^{ab}\t K_{ab}-\t u^i\t \na_i \t K\Big]~.
\label{daneinaajor}
\end{align}
Using the expression of $P^{ab}$ from Eq.\ref{PABTILEIN} in Eq.\ref{daneinaajor}, one finally obtains the desired expression, relating the Lie variation of $P_{ab}$ with the difference between surface and bulk degrees of freedom. 

In the Jordan frame, one can use the same techniques, i.e., define symmetric and anti-symmetric combinations of $\na^a V^b$ (where $V^b$ is any arbitrary vector) as
\begin{align}
S^{ab}[V]=\na^a V^b+\na^b V^a ~~~~~~~~~~~~\textrm{and}~~~~~~~~~~~~ A^{ab}[V]=\na^a V^b-\na^b V^a
\end{align}
Therefore choosing $V^{a}=\phi v^{a}$, and by analogy of the Einstein gravity, we obtain the Noether current in the Jordan frame  to read $J'^a[v]=\na_b A^{ab}[\phi v]$, coinciding with the definition of the Noether potential in Eq.\ref{JAB'}. Note that the Noether current has been arrived at in a purely geometrical manner. Proceeding further, alike the Einstein case, here also we obtain the following identity for the vector field $V^{a}$
\begin{align}
\na_b A^{ab}[V]=2R^a_bV^b+\na_b\Big[\pounds_V g^{ab}-g^{ab}g_{ij}\pounds_Vg^{ij}\Big]~.
\label{AABJOR}
\end{align}
Fixing, $V^a=Nu^a$ in Eq.\ref{AABJOR}, and contracting it with $u_a$, we obtain
\begin{align}
D_a(Na^a)=N\Big[R_{ab}u^au^b+K^{ab}K_{ab}-u^i\na_iK\Big]~,
\label{danaajor}
\end{align}
which yields the desired expression provided in Eq.\ref{idenjor} of the main text.

In arriving at the previous result for Noether current, we have used the fact that $\nabla_{a}\nabla_{b}A^{ab}=0$, for any antisymmetric tensor $A_{ab}$. It also follows for an antisymmetric tensor $A_{ab}$ that, $\nabla_{a}\nabla_{b}(\phi^{n}A^{ab})=0$. Thus, if we fix the arbitrary vector $V^a$ to be $V^a=v^a/\phi^2$, we may write down the following conserved Noether potential, 
\begin{align}
J^{ab}[v]=\frac{\phi^3}{16\pi}A^{ab}\Big[\frac{v}{\phi^2}\Big]~,
\end{align}
which is the Noether potential associated with the Jordan frame action $\mathcal{A}_{\rm J}$, as one can verify through a starightforward simplification of the above expression and its comparison with Eq.\ref{JABJORDAN} in the main text. Thus we have derived all the Noether currents appearing in this work from geometrical identities.  

\section{Noether charge as surface heat content: Inequivalent picture} \label{appenheatcontent}

Given the Noether current in Eq.\ref{JABJORDAN}, associated with the action $\mathcal{A}_{\rm J}$, presented in Eq.\ref{SJ}, we compute $u_aJ^{a}[\xi]$. This turns out to be a total three-derivative:
\begin{eqnarray}
u_aJ^{a}[\xi]=\frac{1}{8\pi}D_b\big(N\phi A^b\big)~;
\qquad
A^b=a^b-D^b\ln\phi=D^b\ln\Big(\frac{N}{\phi}\Big)~.
\end{eqnarray}
To obtain the total Noether charge within a spatial volume, we integrate the density $u_aJ^{a}[\xi]$ with appropriate integration measure, yielding,
\begin{align}
\int_{\R}d^3x\sqrt{h}u_aJ^{a}[\xi]=\frac{1}{8\pi}\int_{\p\R}d^2x\sqrt{\sigma}R_aN\phi A^a~.
\end{align}
Here, $\mathcal{R}$ denotes the $t=$constant hypersurface, $R_a$ is the unit normal to the boundary of the three-volume under consideration. Meaningful thermodynamic interpretation can be obtained, provided we consider the boundary of the three-volume to be defined as $(N/\phi)=\textrm{constant}$ surface within the $t=$constant hypersurface. Then the unit normal $R_{a}$ becomes,
\begin{align}
R_a=\frac{D_a\Big(\frac{N}{\phi}\Big)}{\sqrt{D_i\Big(\frac{N}{\phi}\Big)D^i\Big(\frac{N}{\phi}\Big)}}=\frac{D_a\ln\Big(\frac{N}{\phi}\Big)}{\sqrt{D_i\ln\Big(\frac{N}{\phi}\Big)D^i\ln\Big(\frac{N}{\phi}\Big)}}~,
\end{align}
which implies 
\begin{align}
R_aA^a=\sqrt{A_aA^a}=A~.
\end{align}
Using the above results, one can finally obtain that the Noether charge of the region $\R$ as 
\begin{align}
\int_{\R}d^3x\sqrt{h}u_aJ^{a}[\xi]=\int_{\p\R}d^2x\frac{\phi\sqrt{\sigma}}{4}\frac{NA}{2\pi}=\int_{\p\R}d^2x\,s T_{\rm loc}~, \label{heatcontentineqfinal}
\end{align}
where $s=\phi\sqrt{\sigma}/4$ is the entropy density, which matches with the corresponding result from the action $\mathcal{A}_{\rm J}'$. The redshifted local Unruh temperature becomes, $T_{\rm loc}=(NA/2\pi)$, which depends not only on the acceleration $a^a$, but additional contribution of $\phi$ arises from the non-minimal coupling. However, the above temperature $T_{\rm loc}$ is different from the corresponding one in the Einstein frame even after conformal transformation. Additionally, the two-surface of interest is also not the same. In the Einstein frame, the thermodynamic interpretation is given on $\t N=\textrm{constant}$ surface, which translates into $\sqrt{\phi} N=\textrm{constant}$ surface under conformal transformation, different from the $(N/\phi)=\textrm{constant}$ surface considered here. Nevertheless, both the Noether charges, derived from $\mathcal{A}_{\rm J}$ as well as $\mathcal{A}_{\rm J}'$, corresponding to the action in Eq.\ref{SJ} and Eq.\ref{ACTUAL}, respectively, have the thermodynamic interpretation as the heat content of a suitable two-surface.

\section{Derivation of Eq.\ref{conenction} of the main text} \label{appenconnection}

In this appendix we provide a derivation for Eq.\ref{conenction}, one of the central relations that we have used in the main text. For this purpose, we start from Eq.\ref{a'} and obtain
\begin{align}
N\phi a^a=N\phi a'^a-\frac{1}{2}N D^a\phi~.
\label{appenc1}
\end{align}
Taking three-derivative on both sides of Eq.\ref{appenc1}, presented above, we obtain, 
\begin{align}
D_a(N\phi a^a)=D_a(N\phi a'^a)-\frac{1}{2}Na_aD^a\phi-\frac{1}{2}ND_aD^a\phi~,
\label{appenc2}
\end{align}
where we have used the result, $a_{a}=D_{a}\ln N$. Further, we have the following identity, 
\begin{align}
D_aD^a\phi=\na_a(D^a\phi)-a_aD^a\phi~.
\label{appenc3}
\end{align}
Substituting Eq.\ref{appenc3} in Eq.\ref{appenc2}, we arrive at, 
\begin{align}
D_a(N\phi a^a)-D_a(N\phi a'^a)&=-\frac{1}{2}N\na_a\Big(h^a_i\na^i\phi\big)
\no
\\
&=-\frac{N}{2}\square\phi-\frac{N}{2}\na_a\Big(u^au^i\na_i\phi\Big)
\no
\\
&=-\frac{N}{2}\square\phi-\frac{N}{2}u^a\na_a(u^i\na_i\phi)+\frac{NK}{2}u^i\na_i\phi~.\label{appencfin}
\end{align}
From the above expression, i.e., Eq.\ref{appencfin}, one can obtain Eq.\ref{conenction} of the main text, by simply replacing $Nu^i$ with $\xi^i$.

\section{Gravitational dynamics and holographic equipartition: the inequivalent picture} \label{appenholeq}

In this section we will demonstrate that the gravitational dynamics emerging from the departure of equipartition between surface and bulk degrees of freedom holds for the Jordan frame action $\mathcal{A}_{\rm J}$ as well, but is not equivalent to \gr. Earlier, for the action $\mathcal{A}_{\rm J}$, we found the Noether charge density to read, $16\pi u_aJ^{a}[\xi]=2D_b[N\phi(a^{b}-D^{b}\ln\phi)]=2D_b[N\phi A^{b}]$. We start with the identity in Eq.\ref{idenjor}, and express it as
\begin{align}
2\sqrt{h} D_a\Big(N\phi A^{a}\Big)+2N\sqrt{h}h^{ab}D_{a}D_{b}\phi=2\sqrt{h}N \phi R_{ab} u^a u^b-16\pi \phi h_{ab} \lie P^{ab}~.
\end{align}
Using the following identity: $h^{ab}D_{a}D_{b}\phi=h^{ab}\na_{a}\na_{b}\phi-K(u^{i}\nabla_{i}\phi)$, the above expression becomes,
\begin{align}
2\sqrt{h}D_{a}\Big[\phi N A^a\Big]+2N\sqrt{h}h^{ab}\na_a\na_b\phi-2\sqrt{h}K\lie\phi+16\pi\phi  h_{ab} \lie P^{ab}=2\sqrt{ h} N\phi R_{ab} u^a u^b 
\end{align}
Rearranging the above equation and integrating over a three-dimensional region $\R$, we obtain,
\begin{align}
\int_{\R}d^3x\Big[\phi  h_{ab} \lie P^{ab}&-\frac{1}{8\pi}\sqrt{h}K\lie\phi\Big]=\frac{1}{8\pi}\int_{\R}d^3x\sqrt{h}N\left(\phi R_{ab}-\na_{a}\na_{b}\phi+g_{ab}\square \phi\right)u^{a}u^{b}
\nonumber
\\
&-\frac{1}{8\pi }\int_{\p\R}d^2x\sqrt{\sigma} N\phi R_{a} A^{a}~.
\end{align}
Again, $R_{a}$ is the normal to the surface $(N/\phi)=\textrm{constant}$, acting as the boundary of the spacetime region $\mathcal{R}$ of interest. Thus, it follows that $R_{a}A^{a}=\sqrt{A^{a}A_{a}}$, and hence the local temperature $T_{\rm loc}=(NA/2\pi)$ appears in the above expression. Then we obtain, 
\begin{align}
\frac{1}{8\pi}\int_{\p\R}d^2x\sqrt{\sigma} N\phi R_{a} A^{a}=\frac{1}{4}\k_BT_{\rm avg} n_{\rm sur}~,
\end{align}
where, 
\begin{align}
n_{\rm sur}=\int_{\p\R}d^{2}x\,\phi\sqrt{\sigma}=n_{\rm sur}'~,
\end{align}
and the average temperature $T_{\rm avg}$ is defined as
\begin{align}
T_{\rm avg}=\frac{\int_{\p\R}\sqrt{\sigma}\phi T_{\rm loc}d^2x}{\int_{\p\R}\sqrt{\sigma}\phi d^2x}~. 
\end{align}
We define the Komar energy density associated with the three-volume $\R$ as, 
\begin{align}
\rho_{\rm Komar}=\frac{1}{4\pi}\int_{\R}d^3x\sqrt{h}N\left(\phi R_{ab}-\na_{a}\na_{b}\phi+g_{ab}\square \phi\right)u^{a}u^{b}~,
\end{align}
which on using the field equations in the Jordan frame, namely Eq.\ref{EoMSTalt}, becomes,
\begin{align}
\rho_{\rm Komar}=\int_{\R}d^3x\sqrt{h}2N\left[\bar{T}_{ab}^{\rm (mat)}+\frac{1}{8\pi}\Big\{\frac{\omega}{\phi}\Big(\na_a\phi\na_b\phi\Big)-\frac{1}{2}g_{ab}\square\phi +\frac{V}{2}g_{ab}\Big\}\right]u^{a}u^{b}~.
\end{align}
Following which, the bulk degrees of freedom are defined as
\begin{align}
n_{\rm bulk}=\frac{2}{\k_{\rm B}T_{\rm avg}}\int_{\R}d^3x\sqrt{h}\rho_{\rm Komar}~.
\end{align}
Using the definitions of the bulk and surface degrees of freedom, as presented above, one finally arrives at,
\begin{align}
\int_{\R}d^{3}x\,\Big[\phi  h_{ab} \lie P^{ab}-\frac{1}{8\pi}\sqrt{h}K\lie\phi\Big]=\frac{1}{4}\k_{\rm B}T_{\rm avg}\Big(n_{\rm bulk}-n_{\rm sur}\Big)~,
\end{align}
which is the desired relation for the inequivalent scenario in the Jordan frame. In this case, the thermodynamic parameters (like local temperature and the average temperature) are not equivalent to that of the Einstein frame. In addition, the bulk degrees of freedom are also not equivalent. Nevertheless, in the present case, we can also obtain the holographic equipartition ($n_{\rm bulk}=n_{\rm sur}$) for a static spacetime. In case of departure from the holographic equipartition, it leads to the time evolution of the spacetime. Note that contrary to the relation associated with $\mathcal{A}_{\rm J}'$, the momentum $P_{ab}$ is not conjugate to the spatial metric $h_{ab}$ of the scalar-tensor theory. 



\end{document}